# Visualization of Mesoscopic Conductivity Fluctuations in Amorphous Semiconductor Thin-film Transistors

Jia Yu[1], Yuchen Zhou[2], Xiao Wang[2], Ananth Dodabalapur*[2], Keji Lai*[1]

[1] Department of Physics, University of Texas at Austin, Austin TX 78712, USA

[2] Department of Electrical and Computer Engineering, University of Texas at Austin, Austin TX 78712, USA

* E-mails: ananth.dodabalapur@engr.utexas.edu, kejilai@physics.utexas.edu

## Abstract

Charge transport in amorphous semiconductors is considerably more complicated than process in crystalline materials due to abundant localized states. In addition to device-scale characterization, spatially resolved measurements are important to unveil electronic properties. Here, we report gigahertz conductivity mapping in amorphous indium gallium zinc oxide (a-IGZO) thin-film transistors by microwave impedance microscopy (MIM), which probes conductivity without Schottky barrier's influence. The difference between dc and microwave conductivities reflects the efficacy of the injection barrier in an accumulation-mode transistor. The conductivity exhibits significant nanoscale inhomogeneity in the subthreshold regime, presumably due to trapping and releasing from localized states. The characteristic length scale of local fluctuations, as determined by autocorrelation analysis, is about 200 nm. Using random-barrier model, we can simulate the spatial variation of potential landscape, which underlies the mesoscopic conductivity distribution. Our work provides an intuitive way to understand the charge transport mechanism in amorphous semiconductors at microscopic level.

Amorphous oxide semiconductors are promising material platforms for next-generation transparent and flexible displays since they can be deposited as large-area uniform thin films on plastic substrates at low temperatures[1-5]. Unlike the crystalline counterparts, amorphous materials do not exhibit long-range periodicity in the atomic arrangement[6-8], which usually results in inferior electrical and optical properties. Nevertheless, decades of investigations have led to the discovery of amorphous semiconductors with material qualities sufficient for thin-film transistors (TFTs) applications[5,8]. For instance, in the amorphous indium gallium zinc oxide (a-IGZO) system[2], it is believed that the electronic structures are dominated by the extended spherical $s$ orbitals of heavy metal cations, which are insensitive to the distortion of chemical bonds in disordered structures[2,9,10]. This is in sharp contrast to hydrogenated amorphous silicon (a-Si:H) with strong spatial directivity of the *sp$^3$* orbitals, where the bond-angle distortion leads to significant reduction of carrier mobility[11]. A comprehensive understanding of the charge transport mechanism in a-IGZO is therefore crucial for improving the performance of TFT devices.

The typical mobility of a-IGZO (around 1 ~ 10 cm$^2$/V·s) resides in an intermediate window between that of highly disordered solids (< 0.1 ~ 1 cm$^2$/V·s) and good crystalline semiconductors (> 10 ~ 100 cm$^2$/V·s) [1-5,9,10]. As a result, charge transport in a-IGZO cannot be fully described by either the thermally activated hopping mechanism or the extended-state band model[12,13]. An extended multiple trap and release (MTR) model has been proposed to help explain transport in amorphous oxide semiconductors[14]. In several theoretical studies, the effect of structural disorders in a-IGZO is modeled as a broad distribution of shallow band-tail states and deep traps in the energy gap[15-17]. These localized states give rise to spatial variation of the band edge in the form of random potential barriers with a Gaussian distribution of heights[18]. Electrons moving through the energy landscape are subjected to multiple trapping and releasing processes that determine the transport characteristics of TFT devices[12-18]. On the other hand, experimental investigations of a-IGZO to date have mostly relied on macroscopic transport measurements, which yield little information on the strength and length scale of such potential barriers. In this work, we report the nanoscale imaging of a-IGZO TFT devices by microwave impedance microscopy (MIM)[19-21]. The measured gigahertz (GHz) ac conductivity is higher than the dc conductivity, which can be explained by the contribution from localized electrons. With microwave conductivity measurements, we can probe conductivity without the influence of the Schottky barrier, which is substantial under subthreshold conditions. Additionally, in the subthreshold regime, we observe local conductivity fluctuations with an amplitude of 10 – 20% of the overall values. Autocorrelation analysis of the MIM images indicates that the characteristic length scale of the spatial conductivity variation is ~ 200 nm, i.e., in the mesoscopic regime. The disorder potential landscape can be simulated by using the same length in the random-barrier model. Our work is important for both fundamental understanding and practical application of amorphous semiconductors. For example, conductivity fluctuations on a scale of 200 nm will impact nanoscale a-IGZO and related transistors with channel lengths below 100 nm, that are sought for backend of the line (BEOL) applications[22,23].

The a-IGZO thin films (20 nm in thickness) in this experiment[24] are prepared by RF-sputtering onto heavily doped n-type Si substrates with 90 nm thermal $SiO_2$. The nominal composition of $Ga_2O_3$:$In_2O_3$:$ZnO$ is at a ratio of 1:2:2. The deposition is performed in Ar gas with 7% $O_2$ content and a total pressure of 5 mTorr. The films are then annealed on a hotplate for 1 h at 400 °C to generate oxygen vacancies[25]. The TFT devices are either directly deposited through a shadow mask or lithographically defined after a blanket deposition on the substrate. A layer of 80 – 100 nm Al is evaporated onto the sample and patterned into source/drain contacts by e-beam lithography. Details of the device fabrication steps can be found in the Methods section. In this work, we have studied a total of 5 devices (see Fig. S1 for their transfer characteristics) and the results are consistent among all samples.

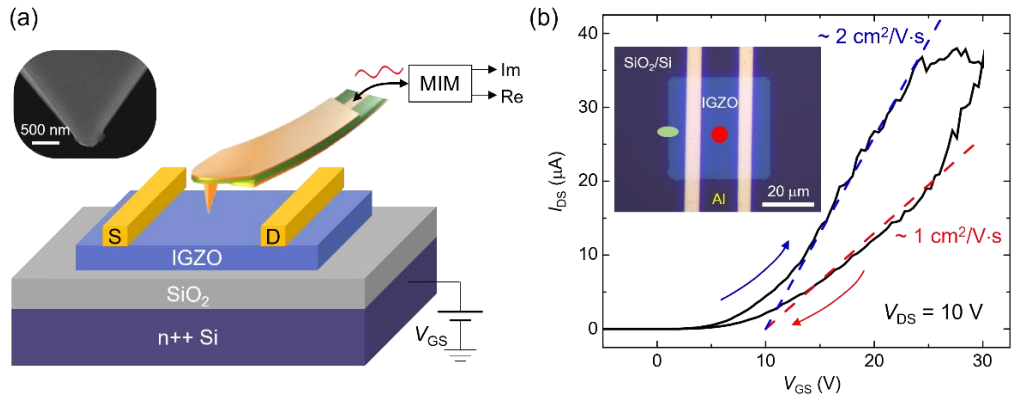

Fig. 1. (a) Schematic of the MIM imaging on a back-gated a-IGZO TFT device. The inset shows the SEM image of the tip apex. (b) Transfer characteristics of a lithographically defined a-IGZO transistor at $V_{DS}$ = 10 V. The field-effect mobility is labeled in the plot. The inset shows an optical image of the device. The approximate scanned areas in Fig. 2a and Fig. 3a are marked by the green oval and red dot, respectively.

Fig. 1a shows the schematic of our experimental setup. The shielded MIM probe (close-up scanning electron microscopy (SEM) image of the tip apex shown in the inset) delivers the excitation signal to the tip apex and the reflected wave is demodulated by microwave electronics[20].

The MIM outputs are proportional to the imaginary (MIM-Im) and real (MIM-Re) parts of the tip-sample admittance, from which the local microwave conductivity can be extracted[19]. The a-IGZO film is patterned into a back-gated TFT structure, as seen in the inset of Fig. 1b. A typical transfer curve of the device at $V_{DS}$ = 10 V and a bias sweep rate of 0.76 V/sec is shown in Fig. 1b. The gate hysteresis between forward and backward sweep is presumably due to the bias stress effect or interfacial defects, which are commonly observed in a-IGZO TFTs[26]. The extracted field-effect mobility of 1 ~ 2 cm$^2$/V·s is also representative for the a-IGZO devices measured in this work.

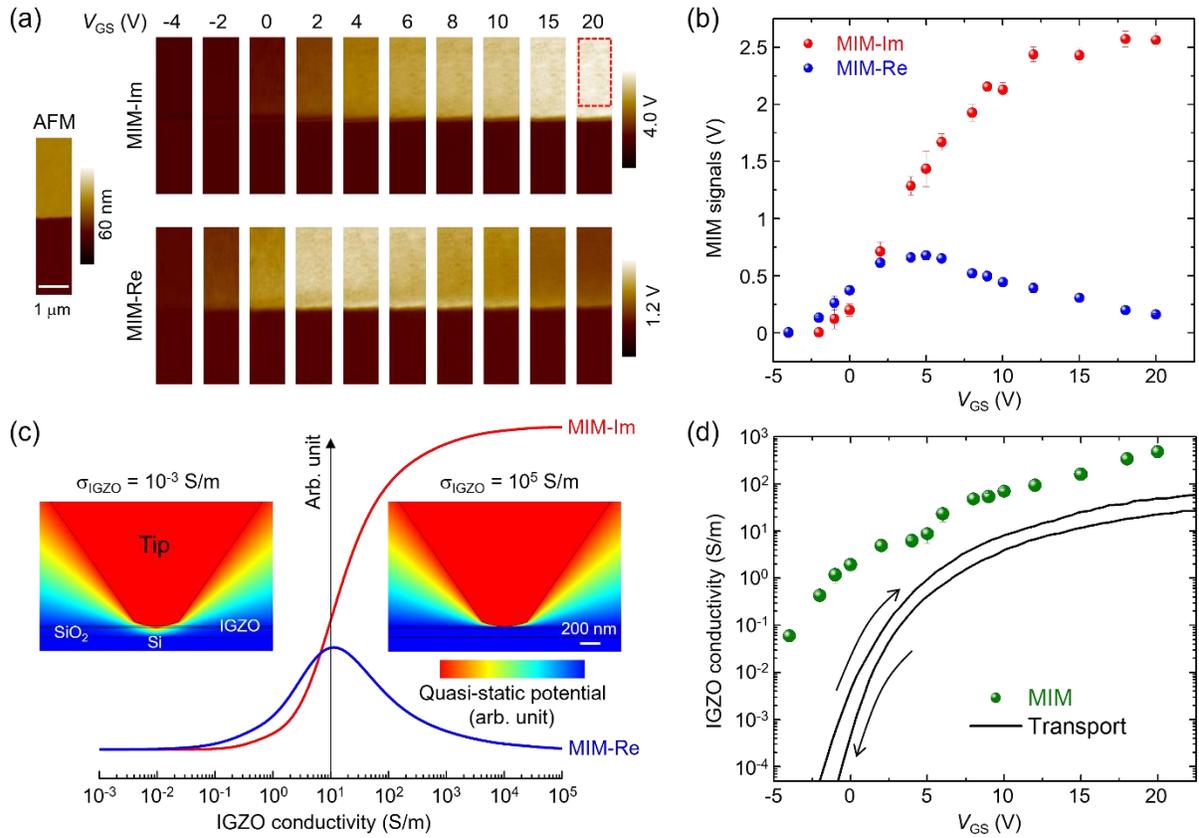

Fig. 2. (a) AFM and MIM images across a step edge of the device in Fig. 1b under various gate biases. The upper part is a-IGZO and the lower part is the SiO$_2$/Si substrate. The dashed box denotes the region to extract the average signals and standard deviations. A complete set of AFM images can be found in Fig. S2. (b) MIM-Im and -Re signals on a-IGZO extracted from the images. (c) Simulated tip-sample admittance as a function of the 1 GHz conductivity in a-IGZO. The insets show the potential distribution under two limits of the ac conductivity. (d) Comparison between microscopic ac conductivity measured by MIM and macroscopic dc conductivity measured by transport.

Fig. 2a shows the 1 GHz MIM images across a step edge of the a-IGZO device under various back-gate voltages ($V_{GS}$). Note that the MIM signals at the step edge are slightly different from that in the central region, presumably due to defects introduced in the etching process. As a result, we only select the center area (dashed box in Fig. 2a) for analysis below. As $V_{GS}$ increases

from –4 V to 20 V, the MIM-Im signals on a-IGZO (contrast to signals on the SiO$_2$/Si substrate) increase monotonically, whereas the MIM-Re signals rise to $V_{GS}$ = 4 V and start to drop afterwards, as plotted in Fig. 2b. To obtain a quantitative understanding of the MIM data, we have performed finite-element analysis (FEA)[19] using the actual tip-sample geometry, as shown in Fig. 2c. By comparing the experimentally measured MIM signals and numerically simulated FEA results, we can extract the microwave ac conductivity $\sigma_{ac}$ as a function of $V_{GS}$, as plotted in Fig. 2d.

For comparison, Fig. 2d also includes the transport dc conductivity $\sigma_{dc}$ calculated from the FET data in Fig. 1b. For $V_{GS}$ > 5 V, $\sigma_{dc}$ is about 10 times lower than $\sigma_{ac}$ and the ratio of $\sigma_{ac}/\sigma_{dc}$ stays roughly constant. At first sight, it is surprising to see the difference between $\sigma_{ac}$ and $\sigma_{dc}$ after the TFT device turns on and the transport is dominated by extended states. The difference between the dc and microwave conductivities in the subthreshold regime can be viewed as a direct measure of how effective the Schottky barrier is in keeping the off-state current (below $10^{-11}$ ~ $10^{-10}$ A) small. In other words, such a large difference implies good off-state characteristics, and the barrier helps keep the current small in the off state and subthreshold regime. A conductivity difference above threshold indicates that the barrier effect persists, to some degree, and manifests as a contact resistance that is generally considered undesirable. The existence of large contact resistance[27-30] can be attributed to contamination during lithography patterning and that the device has relatively short channel length. In contrast, Fig. S3 shows the results from a shadow-mask defined device with much longer contact width, where the contact resistance is small and $\sigma_{ac}/\sigma_{dc}$ indeed approaches 1 at high $V_{GS}$. Since the MIM does not suffer from Ohmic contact and other extrinsic effects[20,21], it is likely that electrical properties of the extended states in a-IGZO are better evaluated by the ac measurement, i.e., the intrinsic mobility of the device in Fig. 2 is ~ 10 cm$^2$/V·s rather than ~ 1 cm$^2$/V·s.

On the other hand, as $V_{GS}$ drops below 5 V, the bias dependence of the $\sigma_{ac}$ measured by MIM is much weaker than that of the $\sigma_{dc}$ measured by transport. This effect can no longer be explained by the influence of contact resistance and can be attributed to the effects of the injection barrier, as described above. For amorphous semiconductors, the significantly larger $\sigma_{ac}$ than the corresponding $\sigma_{dc}$ is well documented in the literature[31-33]. At high (MHz – GHz) excitation frequencies, trapped charges that do not participate in the dc conduction can contribute to ac conduction through hopping between adjacent sites or tunneling across potential barriers. As a result, in the subthreshold regime where transport across macroscopic samples is strongly suppressed, the microscopic ac conductivity may still be appreciable, which is consistent with our observations in Fig. 2d.

The MIM imaging with nanoscale resolution provides further information beyond the average ac conductivity. Since the MIM-Re signals are less susceptible to topographic crosstalk than the MIM-Im counterparts[21], only the former will be presented here. The influence of surface roughness on the MIM signals is discussed in Fig. S4 of the Supplementary Information. Fig. 3a displays selected MIM-Re images in the interior of the a-IGZO device (red dot in Fig. 1b) under different $V_{GS}$. Note that only the relative contrast is presented here since the average signals are automatically removed during the contact-mode MIM imaging[19-21]. The complete set of MIM images and the analysis are included in Fig. S5. When the device is either in the 'off' state or the 'on' state, the images are essentially featureless. On the other hand, the background-removed MIM-Re images display significant inhomogeneity in the subthreshold regime (0 V < $V_{GS}$ < 5 V). Using autocorrelation analysis[34] (Fig. 3a), we can show that the characteristic length scale of such granular features is around 200 nm, as seen in the line profiles in Fig. 3b. Fig. S6 also confirms that the granular patterns are spatially isotropic in nature. We recognize that this dimension is comparable to the tip diameter. Future experiments with sharper tips are needed to verify the length scale of such nonuniformity.

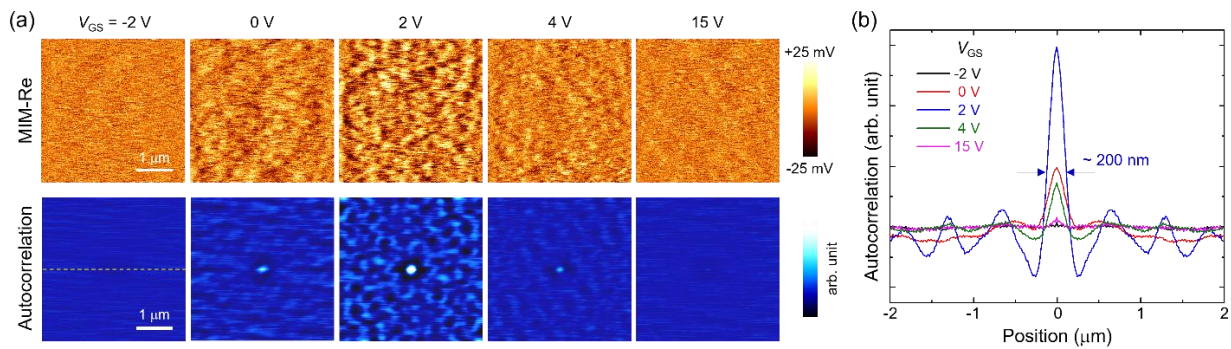

Fig. 3. (a) Top: MIM-Re images on a-IGZO, where the average signals are removed, under different $V_{GS}$. Bottom: Autocorrelation of the corresponding MIM images. (b) Line profiles through the center of the autocorrelation images in (a).

Both the $\sigma_{ac}/\sigma_{dc}$ ratio in Fig. 2d and the local fluctuations of $\sigma_{ac}$ in Fig. 3 can be understood by the vast amount of localized states in amorphous semiconductors. When the a-IGZO TFT is in the 'off' state, the Fermi level is deep inside the conduction band tail[6-8]. Since all charge carriers are trapped in real space, the device is uniformly insulating with low MIM signals and minimal spatial variation. As the Fermi level gradually rises into the subthreshold regime, some electrons may be released from the trap states before becoming localized again by the disorder potential. Such a multiple trapping and releasing process leads to weakly delocalized states around randomly distributed barriers[9,12-18]. The resultant electronic conduction is not adequate to carry transport current across macroscopic distance but can be detected by local ac measurement in the MIM imaging. In Fig. 4a, we convert the MIM data from the shadow-mask device described in Fig. S3 into a conductivity map using the same procedure detailed in Methods and contrast it with the simultaneously taken AFM image (see Fig. S4b for the raw data). With no corresponding topographic features, the $\sigma_{ac}$ map exhibits 10 ~ 20% fluctuation on top of a background conductivity of ~ 10 S/m. Such conductivity fluctuations, which are independent of contact effects, suggest that care must be taken in the design of nanoscale transistors in which conductivity fluctuations can lead to device-to-device nonuniformity. When the transistor size is in the micron scale, which is typical for display transistors, conductivity fluctuations on the scale of 200 nm will likely not matter as much.

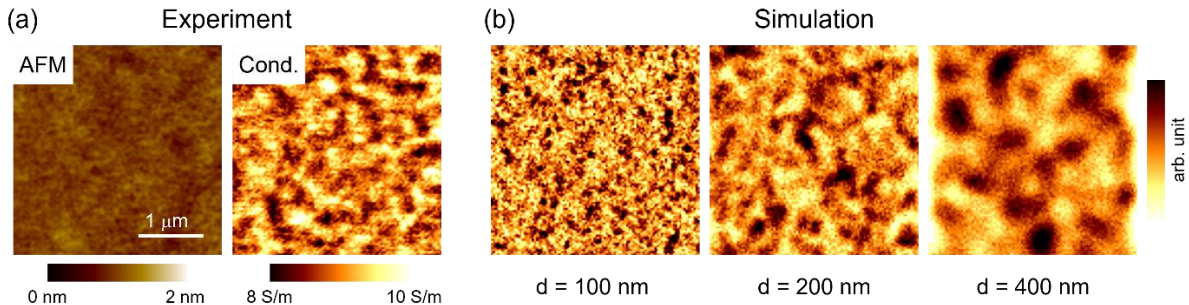

Fig. 4. (a) AFM and ac conductivity images on a shadow-mask-defined a-IGZO device at $V_{GS}$ = 20 V. (b) Simulated potential distribution images with different characteristic lengths of the random barriers. All images have the same dimensions of 3.2 μm × 3.2 μm.

As a final remark, the mesoscopic energy landscape responsible for the conductivity fluctuations can be simulated by the random-barrier model[13,17,26,35]. Here each potential barrier is described by a Gaussian profile $H(x,y) = H_0 \exp\left[-\frac{(x-x_0)^2+(y-y_0)^2}{r^2}\right]$, where $H_0$ is the barrier height (average value of $\mu_1$ and standard deviation of $\Delta_1$), $(x_0, y_0)$ is the coordinate at the center, and $r = d/2$ represents its spatial spread. To mimic the experiments, we also include random noise (average value of 0 and standard deviation of $\Delta_2 = \mu_1/2$) to each pixel in the simulation. For a $D \times D$ frame, the total number of potential barriers $N$ is determined by $(D/r)^2$ such that they can cover the entire grid without significant overlaps. In the numerical modeling, we use $\mu_1 = 100$ meV from a recent analytical report[26] and $\Delta_1 = 25$ meV that is comparable to the thermal energy at room temperature. Fig. 4b shows the simulated potential landscape with $d = 100$ nm, $200$ nm, and $400$ nm. It is obvious that, by matching the width of the Gaussian-like potential profile to the autocorrelation peak width in Fig. 3b, we can reproduce the pattern of conductivity fluctuations in a-IGZO devices. Future experimental and analytical works are needed to extract quantitative information of the disorder potential from the MIM data.

In summary, we report the imaging of GHz conductivity in amorphous IGZO thin-film transistors by microwave impedance microscopy (MIM). The large difference between microwave and dc conductivities below threshold is a reflection of the injection barrier present at the contact that influences the dc current. A large difference below threshold indicated good turn-off characteristics in the device, whereas a small difference above threshold indicates a small contact resistance. Significant local fluctuations of the MIM signals are observed in the subthreshold regime, where a characteristic length scale of ~ 200 nm is determined by autocorrelation analysis. The mesoscopic potential landscape can be simulated by the random-barrier model and the pattern closely mimics the electrical inhomogeneity measured by MIM. Our work provides new insights on the microscopic charge transport mechanism in amorphous semiconductor devices, which are important for improving their performance and expanding their applications.

## Methods

**Amorphous IGZO deposition.** The a-IGZO material was deposited onto a n-type Si substrate with 90 nm of thermal $SiO_2$ through RF-sputtering. The composition of $Ga_2O_3$:$In_2O_3$:$ZnO$ is at a ratio of 1:2:2. The target was ignited at 50 W and slowly ramped up to 150 W at a ramp rate of 5 W/s. A 5-minute deposition was performed with an oxygen percentage of 7% in Ar at a pressure of 5 mTorr. The deposited IGZO was then annealed on a hotplate for 1 h at 400 °C to introduce oxygen vacancies.

**Device fabrication and transport measurements.** The a-IGZO device with well-defined shape shown in the main text was obtained via lithography and etching. The deposited film was first treated by photolithography using AZ5209 as the image reverse resist. The unwanted region was etched away using a 1:6 diluted HCl solution (50 mL HCl:300 mL $H_2O$) for 5 sec and immersed in acetone for 2 min to ensure a clean photoresist liftoff. It then went through standard e-beam lithography process using PMMA as the resist and metal deposition of 80 nm Al contact. Al electrodes are used because the electron affinity of a-IGZO matches with the work function of Al, which facilitates the carrier injection at the metal/semiconductor interface. The substrate was immersed in hot acetone bath inside a glass container for 2 hours at 80 °C for a clean liftoff while avoiding damaging the IGZO material. A cap was used to prevent rapid evaporation of the acetone during the liftoff. The a-IGZO device with larger dimensions shown in the supporting information was achieved by using the shadow masks twice when depositing a-IGZO film and 100 nm Al contact. Neither lithography nor etching was needed. Transport measurements were carried out in the atmosphere using a Keysight 4155C Semiconductor Parameter Analyzer in the dark environment. No appreciable changes in the I-V characteristics were observed after repeated measurements, indicative of the absence of Al diffusion into a-IGZO films.

**Microwave Impedance Microscopy measurements.** The MIM experiments in this work were performed on an AFM platform (ParkAFM XE-70). The customized shielded cantilevers are commercially available from PrimeNano Inc. The tip apex is made of Ti/W, whose work function is close to the electron affinity of a-IGZO, such that the contact potential is negligible. Before each measurement, we first scan a standard sample (e.g., patterned Al dots on a sapphire substrate) with only capacitive contrast. The mixer phase is then adjusted such that the contrast between Al and sapphire only appears in one channel, i.e., the MIM-Im channel. The orthogonal output channel is thus MIM-Re.

**Finite-element analysis.** The commercial software COMSOL 5.4 was used to perform finite-element analysis (FEA), which simulates the real and imaginary parts of the admittance for the specific tip-sample geometry. In this work, we calculated the simulated ratio between MIM-Re and MIM-Im result and compared it with the measured data to determine the ac conductivity. Any circuit-specific factors are cancelled out by using this ratio[36].

**SUPPORTING INFORMATION**

The Supporting Information is available online, including the transfer curves for all devices in this work, complete AFM images for Fig. 2, transport and MIM images of a shadow-mask defined device, topographic crosstalk, complete MIM-Re and autocorrelation data for Fig. 3, and linecuts of autocorrelation images along different directions.

## ACKNOWLEDGMENTS

The MIM work was supported by the U.S. Department of Energy (DOE), Office of Science, Basic Energy Sciences, under Award No. DE-SC0019025. A.D. acknowledges support from the Semiconductor Research Corporation (SRC) task ID # 2962.001, the National Science Foundation under Grant No. NNCI-2025227 as well as the Keck Foundation under Grant No. 26753419. K.L. acknowledges support from the Welch Foundation under Grant No. F-1814. This work was partly done at the Texas Nanofabrication Facility supported by NSF grant #NNCI-1542159.

## AUTHOR CONTRIBUTIONS

K.L. and A.D. conceived the project. Y.Z. and W.X. fabricated the devices and performed the transport measurements. J.Y. carried out MIM experiments, performed data analysis and drafted the manuscript with K.L. All authors have contributed to the manuscript and given approval to the final version of the manuscript.

## NOTES

The authors declare no competing financial interests.

TOC Graphic

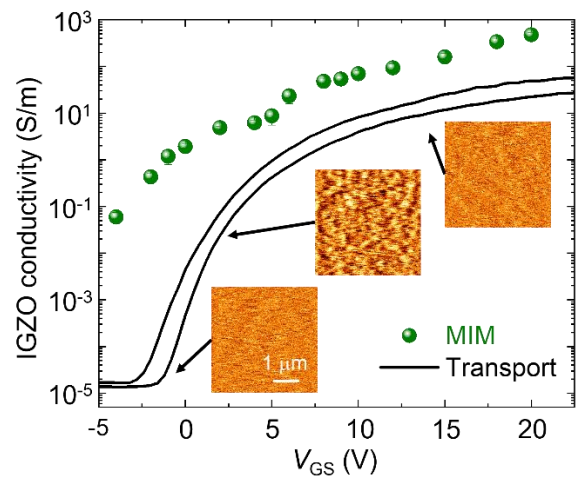